\pdfoutput=1
\documentclass[letterpaper,10pt]{article} 

\usepackage{opticameet3} 

\usepackage{cite}
\usepackage{amsmath,amssymb,amsfonts}
\usepackage{algorithmic}
\usepackage{graphicx}
\usepackage{textcomp}
\usepackage{xcolor}
\usepackage[inline]{enumitem}
\usepackage[linesnumbered,ruled,vlined]{algorithm2e}
\usepackage{svg}
\usepackage{longtable}
\usepackage{float}
\usepackage{caption}
\usepackage{acronym}
\acrodef{OLS}{Open Line System}
\acrodef{ROADM}{Reconfigurable Optical Add-Drop Multiplexer}
\acrodef{BVT}{Bandwidth Variable Transceiver}
\acrodef{OSNR}{Optical Signal to Noise Ratio}
\acrodef{ASE}{Amplified Spontaneous Emission}
\acrodef{NETCONF}{Network Configuration}
\acrodef{NLI}{Non Linear Interference}
\acrodef{BER}{Bit Error Rate}
\acrodef{OSaaS}{Optical Spectrum as a Service}
\acrodef{CFO}{Carrier Frequency Offset}
\acrodef{CDC}{Chromatic Dispersion Compensation}
\acrodef{DGD}{Differential Group Delay}
\acrodef{ORP}{Optical Received Power}
\acrodef{SNR}{Signal to Noise Ratio}
\acrodef{PDL}{Polarization Dependent Loss}
\acrodef{ML}{Machine Learning}
\acrodef{ANN}{Artificial Neural Network}
\acrodef{OCM}{Optical Channel Monitor}
\acrodef{OPM}{Optical Performance Monitoring}
\acrodef{EDFA}{Erbium Doped Fiber Amplifier}
\acrodef{EON}{Elastic Optical Network}
\acrodef{DWDM}{Dense Wavelength Division Multiplexing}
\acrodef{QoT}{Quality of Transmission}
\acrodef{WSS}{Wavelength Selective Switch}
\acrodef{OOK}{On-Off Keying}
\acrodef{PSD}{Power Spectral Density}
\acrodef{SLA}{Service Level Agreement}
\acrodef{XPM}{Cross Phase Modulation}
\acrodef{CPM}{Cross Polarization Modulation}
\acrodef{SOP}{State of Polarization}
\acrodef{PD}{Photodetector}
\acrodef{ILA}{In-Line Amplifier}
\usepackage{setspace}
\usepackage{subcaption}
\usepackage{multirow}
\usepackage[export]{adjustbox}
\usepackage{soul}

\SetCommentSty{mycommfont}

\SetKwInput{KwInput}{Input}                
\SetKwInput{KwOutput}{Output}              
\SetKwFunction{KwFn}{Fn}                   
\usepackage[colorlinks=true,bookmarks=false,citecolor=blue,urlcolor=blue]{hyperref} 
\begin{document}
\vspace{-6mm}
\title{OSNR/GSNR Prediction in Brownfield Links via a DLM-Anchored Hybrid Physics/ML Model}


\vspace{-6mm}
\author{Agastya Raj\textsuperscript{(1)},
Venkata Virajit Garbhapu\textsuperscript{(1)},
Hiroyuki Ishihara\textsuperscript{(3)},
Peyman Pahlevanzadeh\textsuperscript{(1)},
Hideki Nishizawa\textsuperscript{(3)},
Takeo Sasai\textsuperscript{(3)},
Daniel C. Kilper\textsuperscript{(2)},
Marco Ruffini\textsuperscript{(1)}}
\vspace{-1mm}
\address{
\textsuperscript{(1)} School of Computer Science and Statistics, CONNECT, Trinity College Dublin, Ireland\\
\textsuperscript{(2)} School of Engineering, CONNECT, Trinity College Dublin, Ireland\\
\textsuperscript{(3)}Network Innovation Labs., NTT, Inc., Japan}
\vspace{-1mm}
\email{\href{mailto:rajag@tcd.ie}{\textcolor{blue}{rajag@tcd.ie}}}

\vspace{-7mm}

\begin{abstract}
We present a DLM-anchored hybrid physics/ML framework for brownfield optical links that accurately predicts per-channel power, OSNR, and GSNR. Calibrating span/ILA boundaries via DLM yields $\leq$0.39/0.43 dB of OSNR/GSNR errors across single-channel and OSaaS provisioning.
\end{abstract}
\vspace{-3mm}

\section{Introduction}
\vspace{-3mm}
Modern flexible optical networks, including open-line systems (OLS) that enable Optical Spectrum-as-a-Service (OSaaS), require highly accurate quality-of-transmission (QoT) estimation when provisioning new lightpaths in brownfield environments. However, traditional analytical models (GN/EGN) degrade in the field when span loss, connector/WSS insertion loss, or amplifier gain tilt are unknown or drifting, forcing operators to apply multi-dB design margins, reducing efficiency. Machine-learning~(ML)-based component models achieve high per-device accuracy and transferability~\cite{edfa_model}, but degrade in chained networks due to unobserved interstage losses. Hybrid physics/ML methods~\cite{input_refinement} reduce design margins by learning uncertain parameters, but typically require hundreds of measured lightpaths to reach sub-0.2-dB accuracy. More recent work~\cite{zehao_paper} predicts optical signal-to-noise ratio~(OSNR)/general signal-to-noise ratio~(GSNR) with cascaded, pre-trained component models fine-tuned end-to-end, yet still requires extensive per-component and end-to-end measurements across diverse channel configurations (e.g., fully loaded), making it better suited to greenfield than brownfield deployments. A key constraint in brownfield operation is that many link parameters are unknown and live services cannot be disturbed. Addressing this observability gap, Digital Longitudinal Monitoring~(DLM), introduced by~\cite{tanimura_fiber_2020,sasai_digital_2022} provides per-span power profiles from end transceivers, eliminating the need for dedicated inline sensors and accelerating link characterization. However, while DLM exposes the link’s physical state, it does not predict QoT under new operating conditions common in modern, dynamic networks.  Thus, a real-time, high-fidelity QoT prediction solution for disaggregated brownfield networks is still missing.

In this paper, we present a DLM-anchored hybrid physics/ML framework that enables accurate QoT predictions in brownfield links. The framework consists of a chain of component models mirroring the brownfield link, with learnable insertion-loss~(IL) layers at each interface. The chain comprises of analytical Stimulated Raman Scattering (SRS)-based fiber model, static-loss WSS elements, and pre-trained ML EDFA gain/NF models. The ML models were trained on laboratory devices rather than on the target network hardware. We sweep a DLM probe across twelve anchor wavelengths on the brownfield link without disturbing live traffic. The resulting DLM power profiles, together with in-network telemetry, are used to calibrate wavelength-dependent span losses and inline-amplifier (ILA) boundary conditions. Using the same measurements, we jointly fine-tune the EDFA NN models and the IL layers end-to-end, while keeping the core physics modules fixed. Auxiliary OSNR/GSNR models then combine per-channel gains/NFs with ASE (from NF+gain) and NLI from the inter-channel SRS (ISRS)-aware GN/EGN model to produce the final OSNR/GSNR. Once calibrated, the framework predicts per-channel power, OSNR, and GSNR for unseen channel configurations on the same route without additional test traffic. We validate it on a live 175-km mixed-vendor testbed across multiple link settings. Experiments show that DLM-calibrated, pre-trained components achieve 0.20 dB MAE for end-power spectrum prediction and 0.39/0.43 dB MAE for OSNR/GSNR on unseen configurations, with transferability across EDFA gain settings.
\vspace{-4mm}

\section{Experimental Setup}
\vspace{-2mm}

We evaluate the approach on a 175 km brownfield path comprising two Lumentum ROADMs and two ILAs, setup in the OpenIreland Testbed~\cite{edfa_model}, as shown in Fig.~\ref{topology}. The path carries nine C‑band coherent signals from mixed‑vendor transceivers: Cassini, Adtran QuadFlex and Teraflex, at 200 Gb/s on a 50‑GHz grid. Amplifiers operate in constant gain mode, with a per‑channel launch power of 0 dBm at each span. 

\textbf{Data Acquisition:} We use a DLM probe to sweep at twelve random anchor wavelengths across the C-band~(shown in Fig.~1(a)), maintaining$\geq$50-GHz separation from live channels. The DLM probe is a 130-GBd-class, 800 Gbps signal from a commercial muxponder; a least-squares method \cite{sasai_linear_2024} is applied to the received traces to visualize longitudinal power profiles. During the sweep, telemetry is polled every 60s. Specifically, we collect Optical Channel Monitor~(OCM) spectra at the ROADM Booster/Preamp ports on the 50-GHz grid; photodiode (PD) power readings; ILA input/output totals via built-in monitors (ILAs have no OCMs); transceiver performance monitors (received power, BER, Q-factor) for the live channels; and end-of-path OSNR from an OSA on the terminal-ROADM monitor port. GSNR is measured by mapping BER to end-to-end SNR using vendor-provided transceiver characterization curves; when such data are unavailable, nominal back-to-back references can be used, with a constant inverse-SNR penalty removed to isolate link-only GSNR.

\vspace{-3mm}
\section{Modeling Framework} 
\vspace{-3mm}
\begin{figure}
    \centering
    \label{spectrum}
    \includegraphics[width=0.4\linewidth]{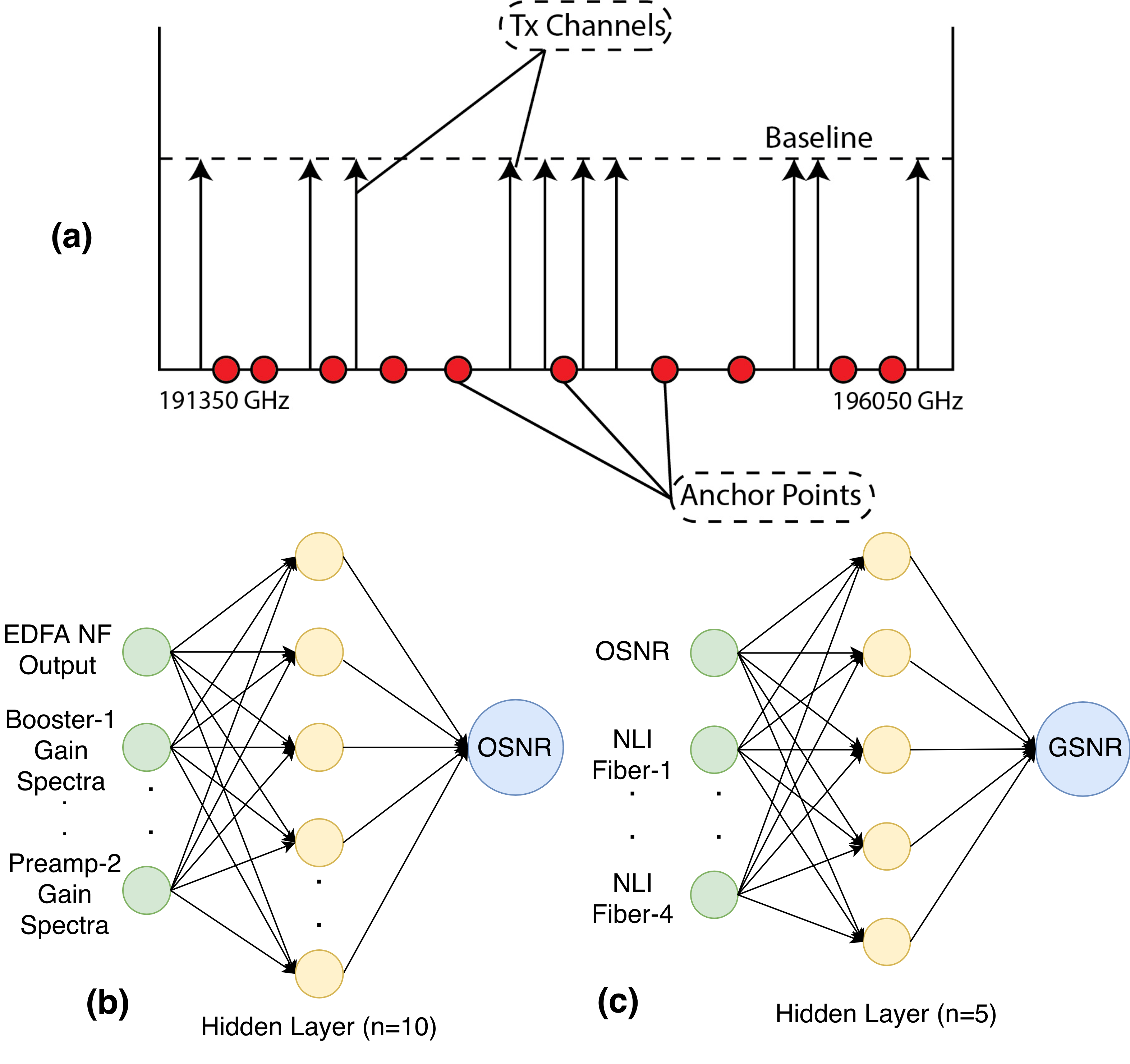}
\includegraphics[width=0.55\linewidth]{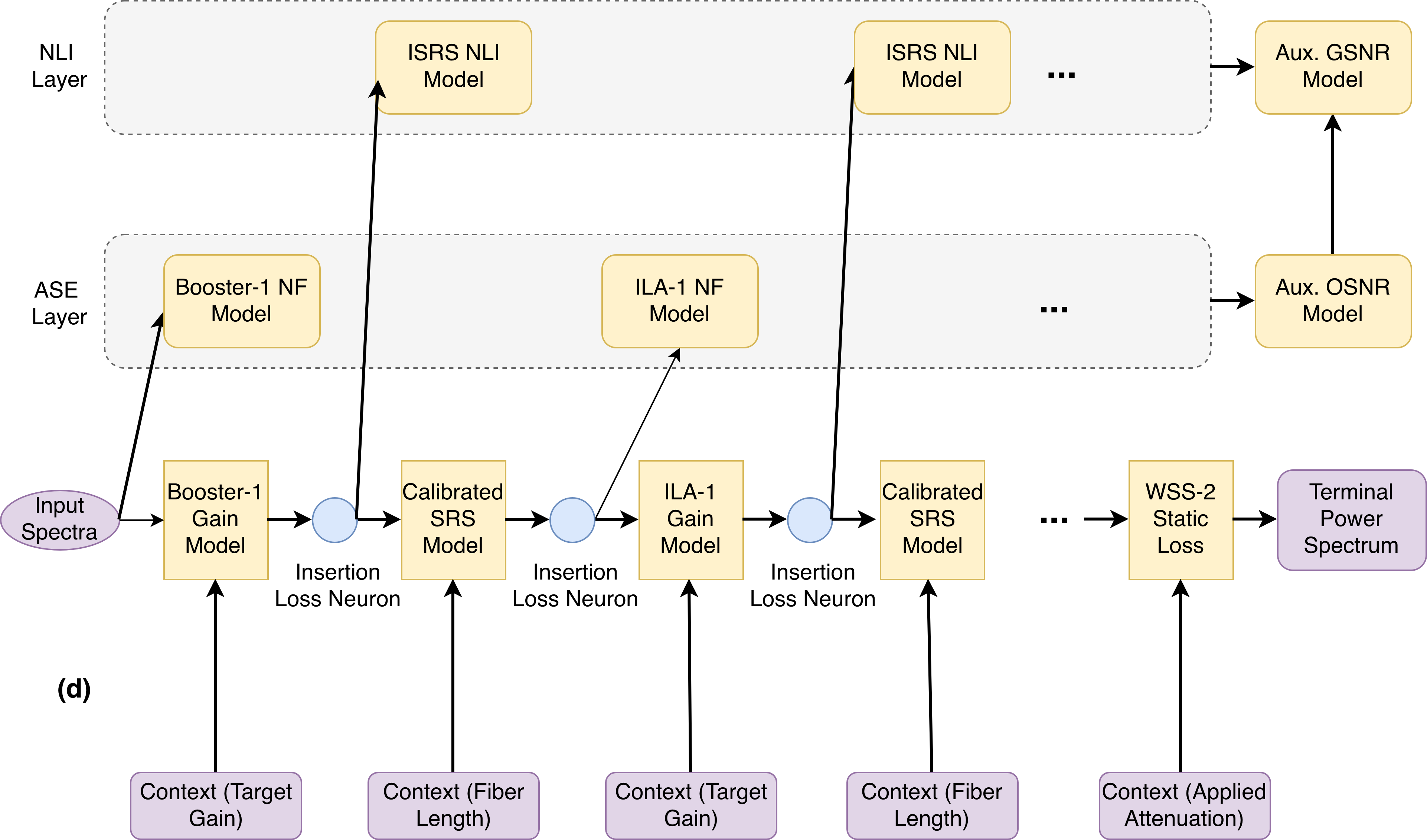}
\vspace{-3mm}
\caption{a) Brownfield spectrum setup: twelve DLM anchor wavelengths are swept across the C-band, (b) Auxiliary OSNR model structure, (c) Auxiliary GSNR model structure, (d) Combined framework: per-channel spectra propagate through chained Booster, SRS-calibrated Fiber, ILA, and WSS static models, with parallel NF and ISRS-NLI branches for OSNR and GSNR prediction.}
    \label{model}
\vspace{-3mm}
    \centering
\includegraphics[width=\linewidth]{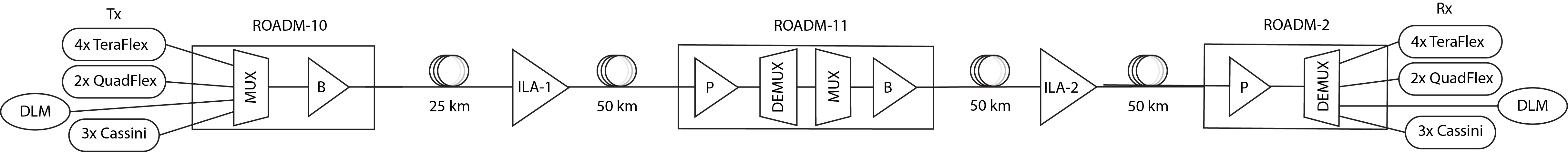}
\vspace{-9mm}
    \label{topology}
    \caption{Experimental topology of the 175 km brownfield optical path. DLM probes are injected and received at the terminal nodes to measure span-wise power evolution along the path.}
    \vspace{-9mm}
\end{figure}

\textbf{Pre-Trained Component Models:} We employ the following models for each network component: (i) Neural Network~(NN) models for amplifier gain and noise figure (booster, preamp, ILAs)~\cite{edfa_model}, trained on lab devices. These models transfer across vendors, eliminating the need to extract or retrain on brownfield hardware, (ii) span-by-span fiber propagation with wavelength-dependent attenuation \(\alpha_s(\lambda)\) [dB/km], dispersion \(D_s(\lambda)\) (slope \(S_s\)), and first-order SRS represented by the coupled Raman power-transfer, and (iii) an ISRS-aware EGN model~\cite{semrau} for nonlinear interference, computing receiver NLI. 

\textbf{Modeling Pipeline:} We construct the framework as an ordered chain mirroring the physical topology, as shown in Fig.~\ref{model}. Each component implements a unified mapping \textit{(input spectrum, context)} \(\rightarrow\) \textit{(output spectrum)}, where context includes target EDFA gains, fiber lengths, and WSS attenuations. Per-channel power spectra propagate stage-by-stage through the component chain, with every block consumes the current spectrum and context and emits an updated spectrum for the next block. Between successive components, we insert learnable single-neuron insertion-loss layers that apply linear corrections to align predicted total power with measured node powers, capturing unmeasured connector and component losses. An auxiliary OSNR model (a single linear layer with masked normalization, masked MAE loss, Gaussian feature-noise, and weight-decay regularization) maps accumulated EDFA noise figures (from boosters + preamps NF models) and per-channel gains to ASE-limited OSNR. We use the ISRS-EGN model to compute NLI under current multi-channel loading using predicted span input spectra. Finally, an auxiliary GSNR model consisting of a single NN layer combines the OSNR prediction with the calculated NLI terms to predict the GSNR. 

\textbf{DLM Calibration: }Transferring the behavior of pre-trained ILA and span models to deployed hardware is challenging, as the intermediate span–ILA–span sections do not expose per-channel power spectra~(ILAs expose only IN/OUT totals), preventing direct fine-tuning of the internal components. Unknown insertion-losses and Raman tilt can be mis-attributed to the ILA, biasing OSNR/GSNR leading to poor generalization. We use per-anchor DLM calibration: (1) segment each longitudinal trace into the two fibers and rescale distances to the known lengths; (2) calibrate that trace to the four node totals from the same run to remove scale/offset drift; (3) extract loss-vs-wavelength and set its mean to the measured node-to-node drop while keeping only the zero-mean DLM tilt. This yields physically plausible, power-closed fiber losses and reconstructed ILA IN/OUT spectra, so the chain cleanly separates fiber effects from the ILA. 

\textbf{Model Training: } For training, we use only the telemetry captured during calibrations (12 DLM sweep configurations). To generalize from limited data while preserving pre-trained behavior, we apply strong L2 weight decay (AdamW) so updates to core EDFA weights are minimal; biases get lighter penalties to correct offset. Additionally, we apply a shape-preserving prior to penalize adjacent-channel slope only when it exceeds a soft threshold $\tau$, and add a curvature penalty on second differences to suppress ripple, evaluated only on active channels. Each EDFA NN is first fit to its own pairs with masked losses over active channels and per-component normalization. We then run teacher-forced chain alignment, feeding measured spectra through the chain and matching predictions at every tap, while learning insertion-loss scalars under a 0.5 dB/connector prior. Finally, we perform end-to-end fine-tuning keeping analytical SRS blocks fixed, using a very low learning rate (1e-5) over 2000 epochs.
\vspace{-4mm}
\section{Inference and Results}  
\vspace{-3mm}
\textbf{Testing Criteria}: We evaluate the model under different channel-provisioning scenarios and link settings. Two provisioning conditions are considered: (a) \textbf{Single‑Channel Provisioning: } one additional 200 Gb/s carrier is activated in one of the vacant 50 GHz slots; and (b) \textbf{Optical Spectrum as a Service (OSaaS) provisioning}, a contiguous 200 GHz block (4 × 50 GHz) with four 200 Gb/s carriers is added. For each provisioning condition, we test three link settings: \textbf{(i) S1:} The EDFA gains are set using the same values from calibration step;  \textbf{(ii) S2:} launch power for all channels reduced by 3 dB relative to S1, and \textbf{(iii) S3:} EDFA gains increased by +3 dB relative to S1/S2. In total, we evaluate 86 single-channel and 15 OSaaS configurations across the three link-gain settings, resulting in 303 measurements, where each measurement represents a complete telemetry capture across all network nodes. It should be noted that the model predictions rely only on the input power spectra at first booster, and component metadata(such as target gain setting); telemetry data is used exclusively for ground-truth comparison and validation, not as inputs during inference. We also evaluate a DLM-free baseline that treats each fiber–ILA–fiber section as a frozen black box: upstream predictions pass through SRS$\rightarrow$pre-trained ILA NN$\rightarrow$SRS, with no DLM calibration or fine-tuning.

\begin{figure}
    \centering
    \includegraphics[width=0.4\linewidth]{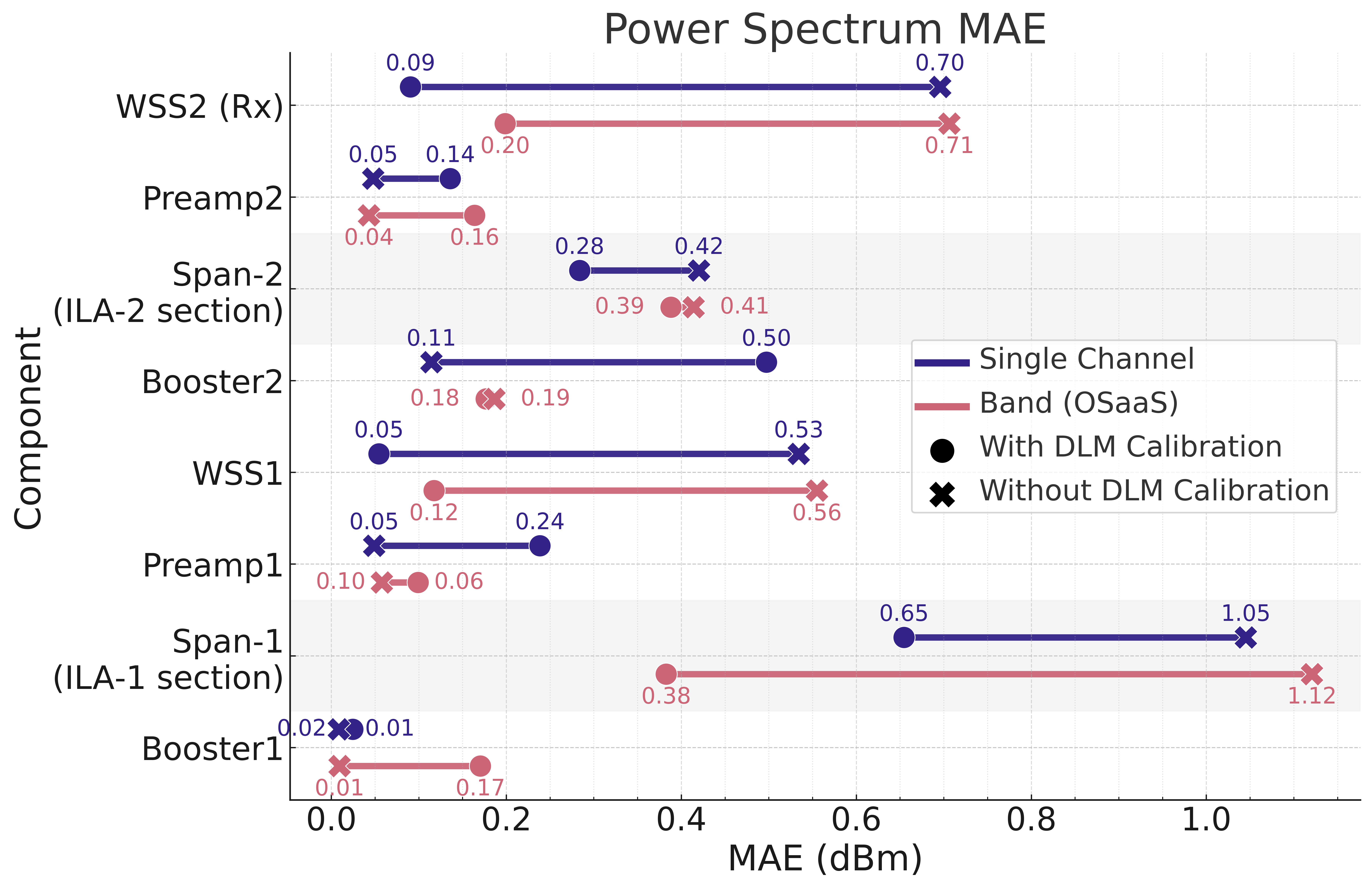}
\includegraphics[width=0.29\linewidth]{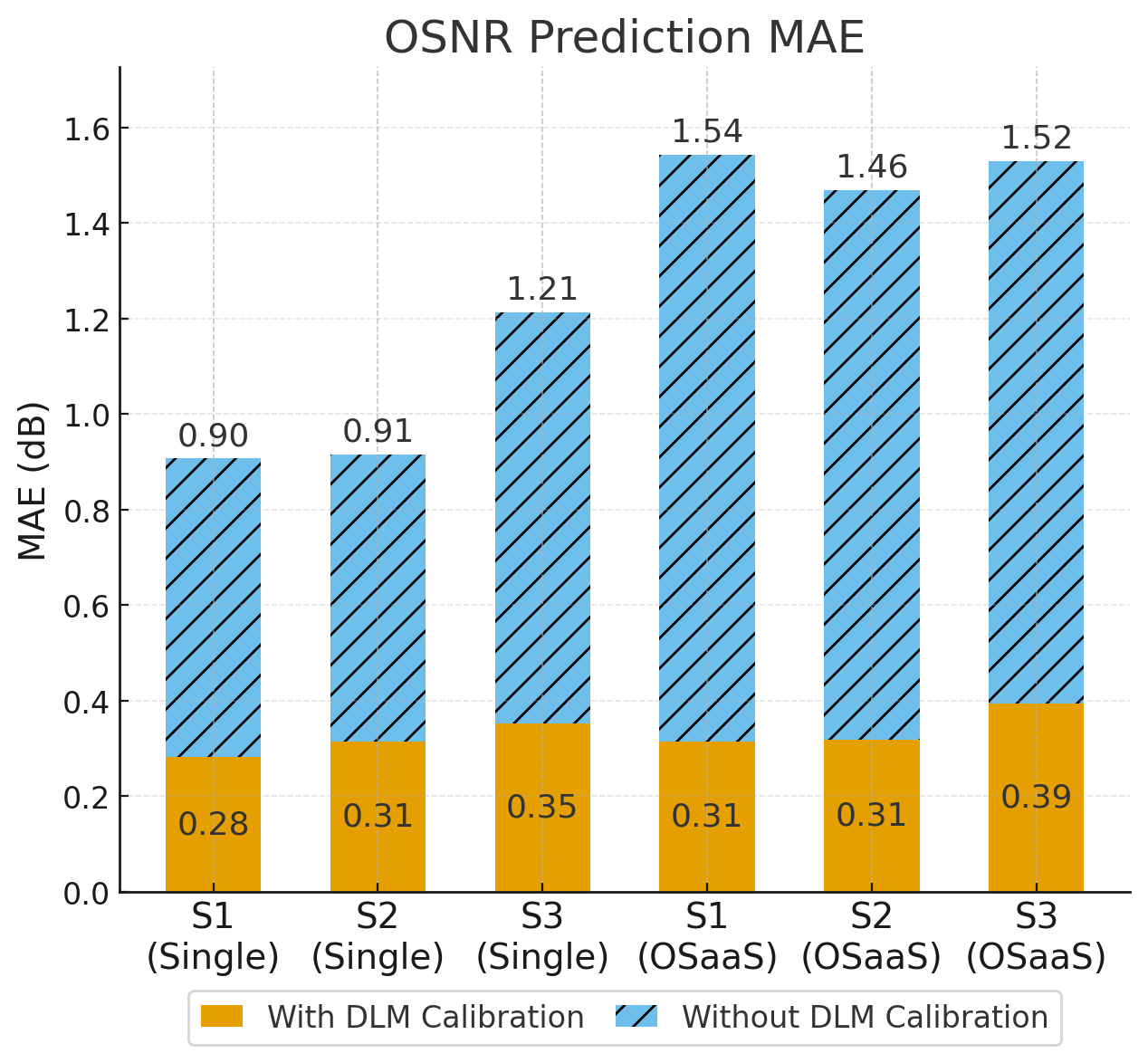}
\vspace{-5mm}
\includegraphics[width=0.29\linewidth]{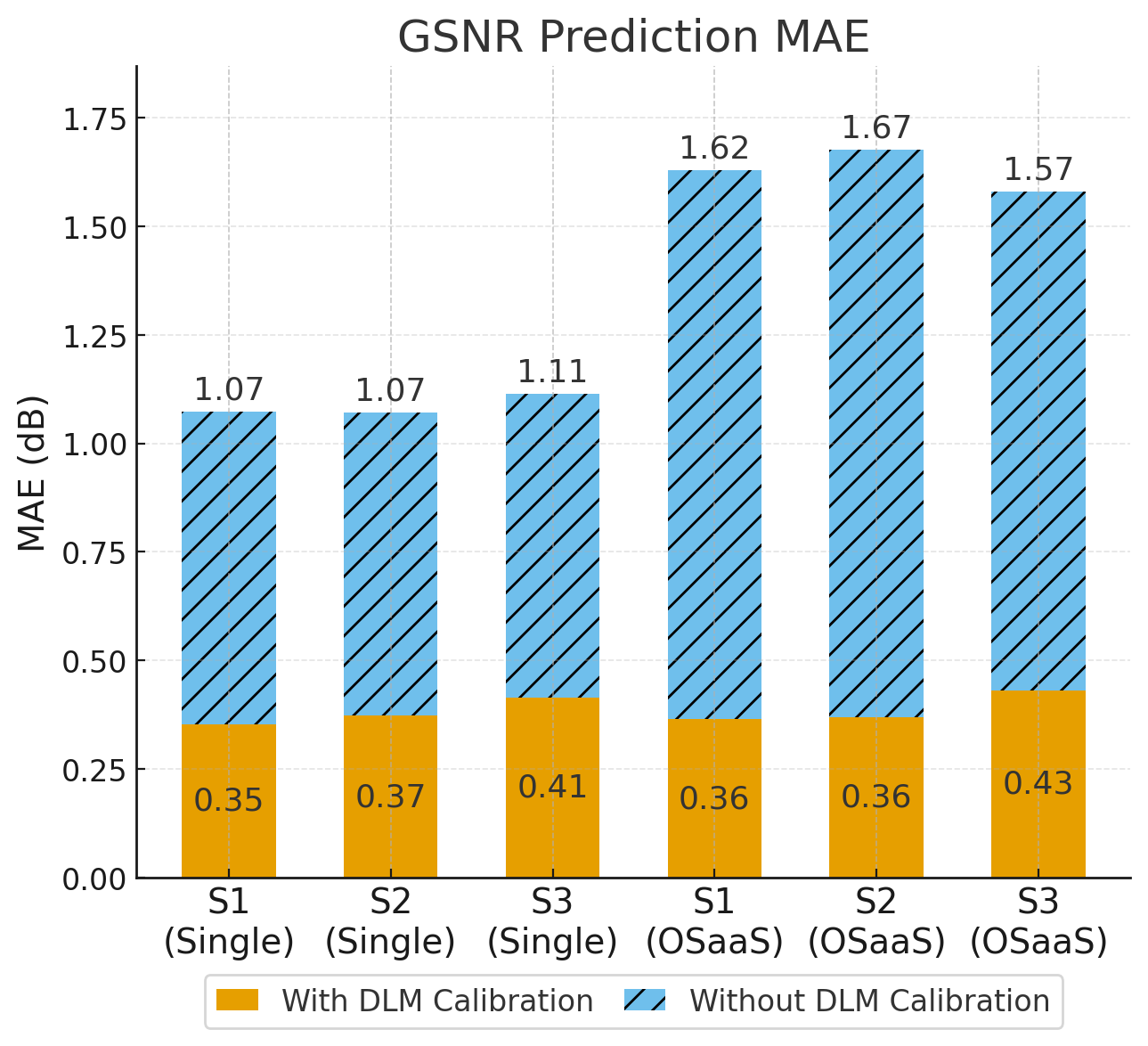}
    \caption{Comparison of model performance under single-channel and OSaaS provisioning scenarios across all link settings.
(a) Mean absolute error (MAE) of power-spectrum predictions for each network component.
(b) MAE of OSNR predictions
(c) MAE of GSNR predictions}
    \label{results}
\vspace{-6mm}
\end{figure}

Fig.~\ref{results}(a) shows the component-wise power-spectrum prediction MAE for single-channel and OSaaS provisioning aggregated across all link settings. DLM calibration consistently reduces component-level errors, with the most significant improvement observed in the ILA sections, where span-level MAE decreases from 1.0–1.1 dB to 0.38–0.65 dB for single-channel and to 0.38 dB for OSaaS provisioning. By calibrating fiber losses and ILA node spectra via DLM, we remove the ambiguity between span loss and ILA gain. The pre-trained ILA then operates as characterized, improving transfer learning and predicting unbiased OSNR/GSNR downstream.

In Fig.~\ref{results}(b), DLM calibration narrows the OSNR MAE to 0.39 dB of error, whereas the baseline without DLM overpredicts the OSNR yielding 0.90-1.54 dB of error. For GSNR prediction, this is further exacerbated, as the MAE for the model without DLM is considerably higher at 1.07-1.67 dB. Comparatively, calibrating with DLM lowers the GSNR predictions to a maximum of 0.43 dB (Fig.~\ref{results}(c)). The strong nonlinear sensitivity of NLI to span-launch power makes post-ILA calibration essential; DLM recovers these hidden boundaries, yielding unbiased GSNR estimates.

\vspace{-5mm}
\section{Conclusions}
\vspace{-3mm}
Using 12 DLM sweeps, our DLM-anchored hybrid physics/ML framework accurately models a 175-km brownfield path, predicting per-channel power, OSNR, and GSNR. Calibrating span losses and node spectra through DLM achieves $\leq$0.39 dB OSNR and $\leq$0.43 dB GSNR errors, enabling test-free provisioning in brownfield links.
\vspace{-1mm}
{\footnotesize
\fontdimen2\font=1pt
\noindent\textbf{Acknowledgments.} Work supported by Research Ireland (RI) grants 12/RC/2276 p2, 18/RI/5721, 13/RC/2077 p2 and 22/FFP-A/10598 and National Institute of Information and Communications Technology (NICT) grant JPJ012368G50201.}

\end{document}